\documentclass[conference, a4paper]{IEEEtran}

\IEEEoverridecommandlockouts
\usepackage{cite}
\usepackage{indentfirst}
\usepackage{graphicx}
\usepackage{svg}
\usepackage{changepage}
\usepackage{stfloats}
\usepackage{amsfonts,amssymb}
\usepackage{bm}
\usepackage{algorithm}
\usepackage{algpseudocode}
\usepackage{amsmath}
\usepackage{amsmath,amssymb,amsfonts}
\usepackage{amsmath, amsfonts, amssymb, amsthm}
\usepackage{times}
\usepackage{url}
\usepackage{cite}
\usepackage{textcomp}
\usepackage{xcolor}
\usepackage{color} 
\usepackage{setspace}
\usepackage{enumitem}
\usepackage{float}
\usepackage{diagbox}
\usepackage[T1]{fontenc}
\usepackage[utf8]{inputenc}
\usepackage{threeparttable}
\usepackage{booktabs}
\usepackage{footnote}
\usepackage{graphicx}
\usepackage{pifont}
\usepackage{subfigure}
\usepackage{mathrsfs}
\usepackage{makecell}
\usepackage{amsmath}
\usepackage[justification=centering]{caption}
\usepackage{lipsum}
\usepackage[numbers,sort&compress]{natbib}
\usepackage{graphicx}
\usepackage{float}
\usepackage{subfigure}
\usepackage{epstopdf}
\usepackage{epsfig}

\usepackage{amsmath}

\newtheorem{definition}{\textbf{Definition}}

\newcommand{\adots}

\begin{document}
\title{Road-Aware Localization With Salient Feature Matching in Heterogeneous Networks}
\author{	
	\IEEEauthorblockN{Lele Cong$^1$, Deshi Li$^1$,  Kaitao Meng$^2$, Shuya Zhu$^1$}
	
	\IEEEauthorblockA{$^1$Electronic Information School, Wuhan University, Wuhan, China.}
	\IEEEauthorblockA{$^2$Department of Electronic and Electrical Engineering, University College London, London, UK.}

	Emails: $^1$\{conglele, dsli, shuyazhu\}@whu.edu.cn, $^2$  kaitao.meng@ucl.ac.uk 
}


\maketitle

\vspace{-10mm}
\begin{abstract}
Vehicle localization is essential for intelligent transportation. However, achieving low-latency vehicle localization without sacrificing precision is challenging. In this paper, we propose a road-aware localization mechanism in heterogeneous networks (HetNet), where distinct features of HetNet signals are extracted for two-spatial-scale position mapping, enabling low-latency positioning with high precision. Specifically, we propose a sequence segmentation method to extract the low-dimensional positioning space on two spatial scales. To represent roads and sub-segments according to HetNet signals, we propose a salient feature extraction method to eliminate redundant features and retain distinct features, thereby reducing feature-matching complexity and improving representation accuracy. Based on the extracted salient features, a two-spatial-scale localization algorithm is designed through salient feature matching, which can achieve low-latency road-aware localization. Furthermore, high-precision positioning is achieved by coordinate mapping based on curve fitting. Simulation results show that our mechanism can provide a low-latency and high-precision positioning service compared to the benchmark schemes.
\end{abstract}


\section{Introduction}
\par
With the development of intelligent transportation systems, vehicle localization is becoming essential in the applications of autonomous driving, vehicle navigation, and traffic management \cite{8827665}. Tremendous demand of low-latency and high-precision vehicle localization exists in road-related localization services, where vehicle localization and navigation accounted for more than 90$\%$ of road-related localization services \cite{requirement}. Especially, to ensure traffic safety, low-latency and high-precision localization is essential for high-speed vehicles. However, the responding time and positioning precision couldn't always be simultaneously fulfilled through the Global Position System (GPS) and the Inertial Navigation System (INS) due to satellite signal loss or long-term cumulative errors \cite{Jian2018MEMS}. With seamless coverage and accessibility, cellular networks provide alternative opportunities for vehicular road localization.
\par
In the existing literature, three typical positioning methods based on cellular networks have been investigated, i.e., the Cell-ID method, the geometric-based method, and the fingerprinting method. Though the delay of Cell-ID methods can be shortened by obtaining a serving cell position, the accuracy relies on the serving cell radius \cite{Zakaria2017PerformanceEO}. In geometric-based methods, location-related information, e.g., received signal strength (RSS) and time of arrival (TOA), of multiple base stations (BSs) is utilized for high-precision position estimations \cite{4510769}. Nevertheless, the localization precision is generally dependent on the explicit line-of-sight (LoS) links \cite{2023Kaitao}. Meanwhile, multiple distance estimation computations between vehicular user equipment (VUE) and BSs result in intolerable response delay \cite{Abbas2018NLOS}. Without relying on LoS links, the radio frequency (RF) signals from multiple transmitters, e.g., Wi-Fi access points (APs) and BSs, can be utilized as fingerprints for localization \cite{Giovanni2018Passive}. However, in wide outdoor environments, fine-grained fingerprint matching is extremely time-consuming to ensure vehicle localization precision. Additionally, the faint decay of millimeter waves according to various environments would greatly affect the localization performances \cite{2017HetNet}. With the development of the fifth generation (5G) beyond and the sixth generation (6G) communication systems, the heterogeneous networks (HetNet) BSs have been widely deployed to improve the network capacity, which provides the potential to reduce localization latency and improve localization precision. Hence, it’s urgent to study a vehicular localization mechanism based on HetNet in complex road environments. 
\par 
Focusing on the fast vehicle localization problem in HetNet, achieving low-latency localization without sacrificing precision is challenging. First, the precision depends on the fine-grained position mapping in wide road environments, resulting in high computation complexity and intolerable delay. Second, representing different roads with sparse signal features is challenging due to complex fluctuations of HetNet signals. To tackle the above problems, we propose a road-aware localization mechanism in HetNet to achieve low-latency vehicular positioning, where distinctive HetNet signal features are extracted to sparsely represent roads and sub-segments. To achieve high-precision localization, coordinates within sub-segments can be localized through curve fitting. The main contributions of this paper are summarized as follows:
\begin{itemize}
	\item We propose a road-aware localization mechanism in HetNet, where salient signal features can be extracted to represent low-dimensional roads and sub-segments, enabling fast localization on roads. 
	\item We propose a sequence segmentation method to extract the low-dimensional positioning space on two-spatial scales. To sparsely represent roads and sub-segments, we propose a salient feature extraction method, which can reduce feature-matching complexity.
	\item We propose a two-spatial-scale vehicular localization algorithm, where the sub-segment posterior probability is obtained by two-spatial-scale feature matching, which can reduce latency and ensure accuracy.
\end{itemize}
\par
The remainder of this paper is organized as follows. Section \ref{model} describes the proposed mechanism and the HetNet signal feature model. In section \ref{Offline}, the HetNet signal feature extraction method is illustrated. The road-aware two-spatial-scale localization algorithm is designed in section \ref{Online}. Section \ref{Experiment Results} provides numerical results to reveal the performance of our proposed mechanism. Section \ref{Conclusions} concludes this paper.
\par
\textit{Notations}: A capital bold-face letter denotes a matrix, e.g., $\boldsymbol{W}$. A vector is denoted by a bold-face lowercase letter, e.g., $\boldsymbol{f}$. The $\mathit{l}_2$ norm of a vector is represented by $\lVert\cdot\rVert_2$.
\section{System Model}\label{model}
\subsection{Road-aware Localization Mechanism}
\begin{figure}[t]
	\centering
	\setlength{\abovecaptionskip}{0.cm}
	\includegraphics[width=8cm]{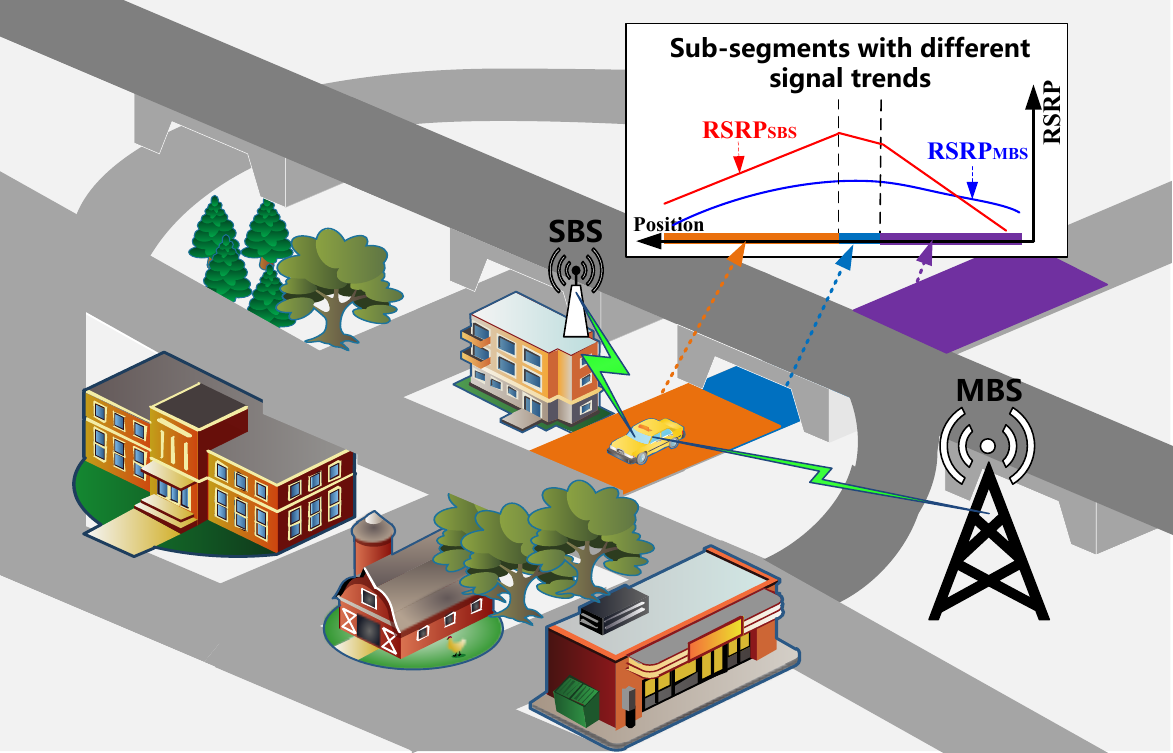}
	\vspace{2mm}
	\caption{Road localization scenario in urban area.}
	\vspace{-2mm}
	\label{figure1}
\end{figure}

\begin{figure}[t]
	\centering
	\setlength{\abovecaptionskip}{0.cm}
	\includegraphics[width=8.5cm]{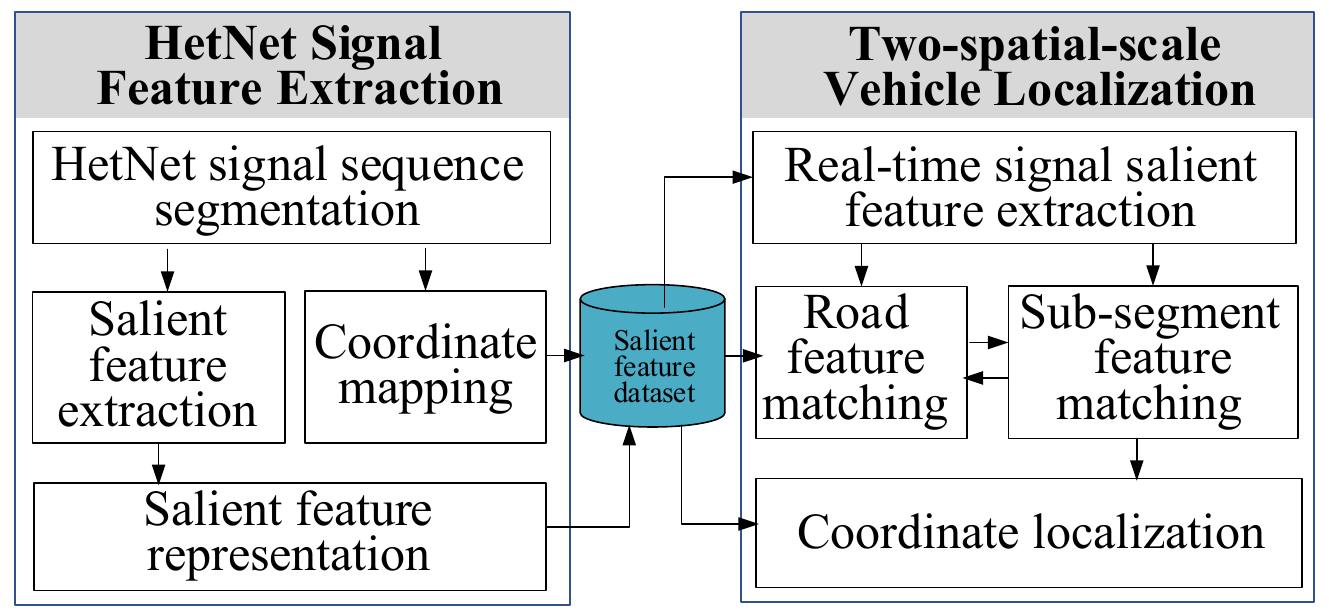}
	\vspace{1mm}
	\caption{Road-aware localization mechanism.}
	\vspace{-6mm}
	\label{figure2}
\end{figure}

The proposed road-aware localization mechanism, depicted in Fig. \ref{figure2}, was designed to achieve low-latency vehicle localization in the complex urban road scenario portrayed in Fig. \ref{figure1}.
Our proposed road-aware localization mechanism can be illustrated by Fig. \ref{figure1} and Fig. \ref{figure2}. Unlike traditional localization based on uniform grids, the wide road environments in Fig. \ref{figure1} can be divided into low-dimensional roads and sub-segments, by analyzing the different fluctuation trends of HetNet signals on roads. Therefore, the positioning computation complexity can be reduced and the road localization precision can be ensured. Highly distinctive signals can be extracted for feature-matching positioning on two-spatial scales, which can further reduce the feature-matching complexity and vehicular positioning delay. To quickly localize vehicular positions on roads, we propose a road-aware localization mechanism with the architecture illustrated in Fig. \ref{figure2}. 
\par
\emph{HetNet signal feature extraction}: To construct the low-dimension positioning space on two-spatial scales, roads are partitioned into sub-segments based on HetNet signal sequence segmentation, which can reduce the positioning computation complexity. To sparsely represent roads and sub-segments, signal features with significant differences can be extracted from two-spatial-scale HetNet signal sequences, thereby reducing feature-matching complexity and achieving low latency. Subsequently, a sparse representation model is designed to represent different salient features. Meanwhile, the reference signal received power (RSRP) values of HetNet signals within sub-segments are mapped to coordinates through curve fitting.   
\par
\emph{Two-spatial-scale vehicle localization}: To achieve vehicle localization with low latency, salient signal features of the real-time HetNet signals should be first extracted. Then, a vehicle localization algorithm is used to match two-spatial-scale salient features, which enables fast road and sub-segment localization. By employing curve fitting, the coordinates within a sub-segment can be localized with high precision.
\subsection{HetNet signal feature model}
\subsubsection{\textbf{HetNet signal sequence}}
HetNet signals received by a vehicle moving on a road can be collected in the form of a signal sequence, which can be leveraged for vehicular road localization. Specifically, a HetNet system is utilized aiming at localizing vehicular positions on roads. The number of HetNet BSs is defined as $K$, i.e., one macro BS (MBS) and $K-1$ small-cell BSs (SBSs). Within a SBS coverage area, the number of roads is defined as $m$, and these roads can be defined as a road space $\mathcal{R}$, i.e., $\mathcal{R}=\left\{\rm{r}_1,\rm{r}_2,\hdots,\rm{r}_\textit{m} \right\}$, where $\rm{r}_\textit{i}$ denotes the $i$th road. The HetNet signal sequence of the road $\rm{r}_\textit{i}$ is defined as a vector $\boldsymbol{\phi}_i$, which is expressed as $\boldsymbol{\phi}_i = \left[\boldsymbol{o}_1, \boldsymbol{o}_2, \hdots, \boldsymbol{o}_L\right]^\top$, where $L$ represents the number of signal sampling positions, and $\boldsymbol{o}_j$ denotes a HetNet signal vector of the $j$th position on the road $\rm{r}_\textit{i}$. The $\boldsymbol{o}_j$ takes the form of $\boldsymbol{o}_j=\left[P_{j,1}, P_{j,2},\hdots, P_{j, K}\right]^\top$, where $P_{j,k}$ denotes the RSRP of the $k$th BS. A position vector is defined as $\boldsymbol{c}_i$ composed by all RSRP sampling coordinates, i.e., $\boldsymbol{c}_i = [[x_1,y_1]^\top,\hdots,[x_L,y_L]^\top]^\top$, where $\left[x_j,y_j\right]^\top$ represents a two-dimensional (2D) coordinate vector of the $j$th position.   
\subsubsection{\textbf{HetNet signal feature}}
The statistical signal features and correlation signal features can be extracted from a HetNet signal sequence to represent corresponding roads, which can reduce computational complexity and improve anti-noise ability. Specifically, for a road-scale HetNet signal sequence $\boldsymbol{\phi}_i$, where $i \in \left[1,m\right]$, the gradient feature model \cite{6821742} takes the form:
\begin{equation}\label{g}
    \setlength\abovedisplayskip{3pt}
	g_{_{j,k}}=  \frac{P_{j+1,k}-P_{j,k}}{\left(\left(x_{j+1}-x_j\right)^2+\left(y_{j+1}-y_j\right)^2\right)^{\frac{1}{2}}}
	,
	\vspace{-1mm}
\end{equation}
where $g_{_{j,k}}$ represents the $k$th BS's signal gradient between the $j$th and the ($j+1$)th positions on the road $\rm{r}_\textit{i}$. Then, the mean signal gradient of the $k$th BS on this road is denoted by  $\overline{g}_k$. To capture correlations between two BS signals, the difference feature model \cite{5462020} is given by:
\begin{equation}\label{dif}
\setlength\abovedisplayskip{3pt}	
	\Delta_{j,k}= P_{j,k}-P_{j,1} ,
	\vspace{-1mm}
\end{equation}
where $\Delta_{j,k}$ denotes the signal difference between the $k$th SBS and the MBS on the $j$th position. Then, the mean difference feature of all positions on this road can be expressed as $\overline{\Delta}_k$. The variance feature ${\sigma}_k$, the mean value feature ${\mu}_k$, and the signal range feature $\chi_k$ of $\boldsymbol{\phi}_i$ can be extracted through general statistical methods for road representations. Hence, a feature set extracted from a road-scale HetNet signal sequence can be defined as $\boldsymbol{e}_{\rm{r}_\textit{i}}$, which is expressed as follows:  
\begin{equation}\label{featureset}
\setlength\abovedisplayskip{3pt}
\setlength\belowdisplayskip{3pt}
	\begin{split}
	\boldsymbol{e}_{\rm{r}_\textit{i}}= \left[\overline{g}_1,\hdots, \overline{g}_K, {\mu}_1,\hdots, {\mu}_K,{\sigma}_1,\hdots, \right. \\	
	 \left.  {\sigma}_K, \overline{\Delta}_1,\hdots, \overline{\Delta}_K, \chi_1,\hdots,\chi_K \right]^\top.
	 \end{split}
\end{equation}	
The feature number of the $\boldsymbol{e}_{\rm{r}_\textit{i}}$ is $qK$, where $q$ represents the number of feature kinds. To reduce localization latency and localization errors, distinct signal features
in $\boldsymbol{e}_{\rm{r}_\textit{i}}$ can be extracted to represent positions on two-spatial scales, which enable low-complexity feature matching and fast vehicle localization.
\section{HetNet Signal Feature Extraction}
\label{Offline}
To represent different spatial-scale roads with accurate signal features, sequence segmentation of HetNet signals and salient signal feature extraction are essential. To establish a fine-grained positioning space, roads should be divided into multiple sub-segments. By utilizing the different fluctuation trends of HetNet signals on roads, the HetNet signal sequences can be segmented into sub-sequences. Then, the corresponding sub-sequences can be used to determine the sub-segments. The definition of a sub-segment is given as follows,
\begin{definition}[\textbf{Sub-segment}]
	\vspace{-2mm}
	A sub-segment is a road segment that possesses similar signal features within itself and different features compared with other sub-segments. A sub-segment position is denoted by a 2D coordinate of the middle position within this sub-segment.
	\vspace{-2mm}
\end{definition}
To accurately distinguish roads and sub-segments with sparse features, distinct signal features should be extracted for position representation, which can reduce the feature-matching complexity and improve representation accuracy. Hence, the salient features are defined as follows, 
\begin{definition}[\textbf{Salient features}]
	\vspace{-2mm}
	The HetNet signal features that exist significant distinctions between roads, and between sub-segments are defined as salient features.
	\vspace{-2mm}       
\end{definition}
However, the segmentation granularity is hard to keep the same value, and extracting the sparse features to represent different roads is challenging. To partition a road-scale HetNet signal sequence, we propose a sequence segmentation method, where all signal gradients are utilized for efficient sequence segmentation. To extract distinctive signal features, we propose a salient feature extraction method through maximizing information gain. Meanwhile, to represent two-spatial-scale salient features, a sparse representation model is designed. The proposed sequence segmentation method, salient feature extraction method, and sparse representation model are explained in the following.
\subsection{HetNet Signal Sequence Segmentation}\label{segprocess}
To partition roads into sub-segments with different signal gradients, we propose a sequence segmentation method by detecting the singular points of a road-scale HetNet signal sequence, which can ensure segmentation accuracy under unequal partitioning granularity. Specifically, a singular point is defined as a HetNet signal vector of a HetNet signal sequence, where the signal gradient of at least one BS changes on this singular point. For a road-scale HetNet signal sequence $\boldsymbol{\phi}_i$, the position index set of all singular points is denoted by $\mathcal{B}=\left\{ b_1, b_2,..., b_{{n_i}-1} \right\} $, where $b_l$ ($b_l \in \left[1, L\right]$) is the position index of the $l$th singular point in $\boldsymbol{\phi}_i$. The $L$ denotes the number of signal sampling positions of $\boldsymbol{\phi}_i$, and the number of singular points is $n_i-1$, which would be different when the road changes. To obtain sub-segment-scale HetNet sequences with different gradients from $\boldsymbol{\phi}_i$, the sequence segmentation method can be designed by extracting the optimal position index set of singular points as follows: 	
\vspace{-1mm}
\begin{equation}\label{segproblem}
	\mathcal{B^\ast} = \mathop{\arg\min}\limits_{\mathcal{B}} 
		{\sum\limits_{k = 0}^{K} \sum\limits_{j = {b_l}}^{b_{l+1}}} \frac{\left(g_{_{j,k}}- \frac{1}{b_{l+1}-b_l}\sum\limits_{j = b_l}^{b_{l+1}-1} g_{_{j,k}} \right)^2}{b_{l+1}-b_l} 
\end{equation}
where $g_{_{j,k}}$ denotes the signal gradient of the $k$th BS on the $j$th position extracted by the gradient feature model in Eq. (\ref{g}), and $K$ is the number of HetNet BSs. Extracting the $\mathcal{B^\ast}$ is highly non-trivial since accurate singular points are used to partition the HetNet signal sequences, which can improve the accuracy of feature extraction and sub-segment representation. To extract the optimal position index set $\mathcal{B^\ast}$ within the HetNet signal sequence $\boldsymbol{\phi}_i$, the Bottom-up algorithm \cite{TRUONG2020} is utilized. Based on two adjacent position indexes of the extracted singular points, a road-scale HetNet signal sequence $\boldsymbol{\phi}_i$ can be partitioned into multiple sub-segment-scale HetNet RSRP sequences, i.e., $\hat{\boldsymbol{\phi}_i}=\left\{ \boldsymbol{\psi}_1,\hdots,\boldsymbol{\psi}_{n_i} \right\}$, where  $\boldsymbol{\psi}_l=\left\{\boldsymbol{o}_{b_l},\hdots,\boldsymbol{o}_{b_{l+1}}\right\}$, and $n_i$ denotes the number of partitioned sub-segment-scale HetNet RSRP sequences. By adopting the position indexes in $\mathcal{B^\ast}$, sub-segments within a road $\rm{r}_\textit{i}$ can be extracted and defined as a sub-segment space $\mathcal{S}_i$, i.e., $\mathcal{S}_i=\left\{ {\rm{s}_1},{\rm{s}_2},\hdots,{\rm{s}}_{n_i} \right\}$, where $\rm{s}_\textit{l}$ denotes the $l$th sub-segment in the road $\rm{r}_\textit{i}$ corresponding to the sub-segment-scale HetNet signal sequence $\boldsymbol{\psi}_l$ in $\hat{\boldsymbol{\phi}_i}$.   
\setlength{\belowdisplayskip}{1pt}
\subsection{Salient Signal feature Extraction}
Due to high-dimensional signals within two-spatial-scale HetNet signal sequences, distinct signal features should be extracted to sparsely represent the roads and sub-segments. Hence, we propose a salient feature extraction method that orients to each road and each sub-segment, which can reduce feature dimension and improve road representation accuracy. Specifically, a feature set vector is extracted from the road-scale signal sequence $\boldsymbol{\phi}_i$ to select salient features representing the road $\rm{r}_\textit{i}$. This feature set vector is denoted by $\boldsymbol{e}_{\rm{r}_\textit{i}}$ as Eq. (\ref{featureset}). The salient features of the road $\rm{r}_\textit{i}$ is expressed as a vector $\boldsymbol{f}_{\rm{r}_\textit{i}}$. To extract the optimal $\boldsymbol{f}_{\rm{r}_\textit{i}}^\ast$ from $\boldsymbol{e}_{\rm{r}_\textit{i}}$ with the maximum information gain, a salient feature extraction method is designed as follows:  
\begin{equation}
	\setlength\abovedisplayskip{2pt}
	\setlength\belowdisplayskip{2pt}
	\boldsymbol{f}_{\rm{r}_\textit{i}}^\ast =\mathop{\arg\max}_{\boldsymbol{f}_{\rm{r}_\textit{i}} \subseteq \boldsymbol{e}_{\rm{r}_\textit{i}}}
	H(\rm{r}_\textit{i})-H\left({\rm{r}_\textit{i}} \mid {\boldsymbol{f}_{\rm{r}_\textit{i}}}\right)
\end{equation}
where the $H(\rm{r}_\textit{i})$ denotes the road information entropy of RSRP values in $\boldsymbol{\phi}_i$, and the $H\left({\rm{r}_\textit{i}} \mid {\boldsymbol{f}_{\rm{r}_\textit{i}}}\right)$ represents the road conditional entropy of the salient feature vector $\boldsymbol{f}_{\rm{r}_\textit{i}}$. The optimal salient feature vector is a subset of the feature set $\boldsymbol{e}_{\rm{r}_\textit{i}}$ corresponding to the maximum information gain, which is extracted by exhaustive search. The calculations of information entropy and conditional entropy are respectively based on the probability of RSRP values and salient features, which are not described in detail here. 
\vspace{-0.1mm}
\par
Due to different feature kinds among the extracted road-scale salient features, a sparse representation model is designed to represent salient feature vectors as follows:
\begin{equation}\label{1}
	\setlength{\abovedisplayskip}{2pt}
	\boldsymbol{f}_{\rm{r}_\textit{i}}={\boldsymbol{W}_{\rm{r}_\textit{i}}}{\boldsymbol{e}_{\rm{r}_\textit{i}}},
	\vspace{1mm}
\end{equation}
where the $\boldsymbol{W}_{\rm{r}_\textit{i}} \in \mathbb{R}^{N \times N}$ represents a feature selection matrix which is denoted by a $N \times N$ diagonal matrix, and $N$ represents the feature number of the feature set vector $\boldsymbol{e}_{\rm{r}_\textit{i}}$, i.e, $qK$. The subscript of the  $\boldsymbol{W}_{\rm{r}_\textit{i}}$ denotes that this feature extraction matrix corresponds to the road $\rm{r}_\textit{i}$. The $w_{kk} (k \in \left[1,qK\right])$ represents the selection coefficient of the $k$th feature, which is defined as a binary variable, i.e., $w_{kk} = 1$ if the $k$th feature of $\boldsymbol{e}_{\rm{r}_\textit{i}}$ is one feature of the extracted salient features. Otherwise, $w_{kk} = 0$. Since the salient feature vector of a road can be represented by multiplying its feature selection matrix and its feature set vector, a vector composed of all road-scale salient features is defined as $\boldsymbol{f}_{\rm{r}}$, which is expressed as follows:
\begin{equation}\label{F_r}
	\setlength{\abovedisplayskip}{2pt}
	\boldsymbol{f}_{\rm{r}}=\left[\boldsymbol{f}_{\rm{r}_1},...,\boldsymbol{f}_{\rm{r}_m}\right],
	\vspace{1mm}
\end{equation}
where $\boldsymbol{f}_{\rm{r}_\textit{i}}$ denotes the salient feature vector of the road $\rm{r}_\textit{i}$, and $\boldsymbol{f}_{\rm{r}_\textit{i}}$ is sparsely represented by ${\boldsymbol{W}_{\rm{r}_\textit{i}}}{\boldsymbol{e}_{\rm{r}_\textit{i}}}$. 
\par
To improve the representation precision of sub-segments, salient features of sub-segments should be extracted as well, since similar gradients may exist in some sub-segments which are not adjacent. To this end, the salient feature vector $\boldsymbol{f}_{\rm{s}_\textit{j}}$ of the $j$th sub-segment $\rm{s}_\textit{j}$ on the road $\rm{r}_\textit{i}$ can be extracted from corresponding feature set vector $\boldsymbol{e}_{\rm{s}_\textit{j}}$ through the proposed salient feature extraction method. Based on the proposed sparse representation model, the extracted $\boldsymbol{f}_{\rm{s}_\textit{j}}$ can be expressed as $\boldsymbol{W}_{\rm{s}_\textit{j}}\boldsymbol{e}_{\rm{s}_\textit{j}}$, where $\boldsymbol{W}_{\rm{s}_\textit{j}} \in \mathbb{R}^{N \times N}$ denotes the feature selection matrix of the $j$th sub-segment $\rm{s}_\textit{j}$. Then, a vector composed of all sub-segment-scale salient features is defined as $\boldsymbol{f}_{\rm{s}}$, taking the form of:
\begin{equation}\label{F_s}
	\setlength{\abovedisplayskip}{2pt}
	\boldsymbol{f}_{\rm{s}}=\left[\boldsymbol{f}_{\rm{s}_1},...,\boldsymbol{f}_{{\rm{s}}_{n_i}} \right],
	\vspace{-0.5mm}
\end{equation}
where $\boldsymbol{f}_{\rm{s}_\textit{j}}$  is sparsely represented by ${\boldsymbol{W}_{\rm{s}_\textit{j}}}{\boldsymbol{e}_{\rm{s}_\textit{j}}}$, and $n_i$ is the number of the sub-segments within the road $\rm{r}_\textit{i}$. 
\section{Two-spatial-scale Vehicle Localization}
\label{Online}
To represent the estimated positions by signal features, the sparse representation model is used to extract vehicular salient feature vectors from the real-time HetNet signal sequence. Based on the extracted two-spatial-scale salient feature vectors, a two-spatial-scale localization algorithm is designed, where the sub-segment posterior probability is utilized through salient feature matching, which can enable low-latency road and sub-segment localization. Moreover, the vehicular coordinates within a sub-segment can be localized through curve fitting.
\subsection{Real-time Salient Signal Feature Extraction}
To effectively extract vehicular salient feature vectors, the vehicular HetNet signal sequence is segmented by singular points extracted in \ref{segprocess} to obtain the real-time HetNet signal sequence. Based on this real-time HetNet signal sequence, the vehicular feature set vector $\boldsymbol{e}_u$ can be first extracted with the same model in Eq. (\ref{featureset}). Then, the vehicular two-spatial-scale salient feature vectors, i.e., $\hat{\boldsymbol{f}}_{\rm{r}_\textit{i}}$ and $\hat{\boldsymbol{f}}_{\rm{s}_\textit{l}}$, can be extracted by multiplying the vehicular feature set vector and the feature selection matrix of the matched road $\rm{r}_\textit{i}$ and sub-segment $\rm{s}_\textit{l}$. 
\subsection{Two-spatial-scale Feature matching}
Due to low computation complexity and high localization precision, the posterior probability of a sub-segment fusing geographical prior information is utilized for vehicular localization by matching two-spatial-scale salient features. Based on the extracted two-spatial-scale feature vectors $\boldsymbol{f}_{\rm{r}}$ and $\boldsymbol{f}_{\rm{s}}$, the vehicular posterior probability of a sub-segment can be obtained by applying the Bayes’ rule as follows: 
\begin{equation} \label{postpro}
	\setlength{\abovedisplayskip}{2pt}
	\hspace{-2mm}
	Pr\left(\rm{s}_\textit{l}\mid \hat{\boldsymbol{f}}_{\rm{s}_\textit{l}} \right) = \frac{Pr\left(\hat{\boldsymbol{f}}_{\rm{r}_\textit{i}} \mid {\rm{r}_\textit{i}}\right)
		Pr\left(\rm{s}_\textit{l} \mid \rm{r}_\textit{i} \right)
		Pr\left(\hat{\boldsymbol{f}}_{\rm{s}_\textit{l}} \mid \rm{s}_\textit{l} \right)}
	{\sum\limits_{q = 1}^{n} Pr\left(\hat{\boldsymbol{f}}_{\rm{s}_\textit{q}} \mid {\rm{s}_\textit{q}} \right)Pr\left({\rm{s}_\textit{q}}\right)}
	\vspace{1mm}
\end{equation}
where the $Pr\left(\hat{\boldsymbol{f}}_{\rm{r}_\textit{i}} \mid {\rm{r}_\textit{i}}\right)$ denotes the matching probability of road-scale salient features, $Pr\left(\hat{\boldsymbol{f}}_{\rm{s}_\textit{l}} \mid \rm{s}_\textit{l} \right)$ and $Pr\left(\hat{\boldsymbol{f}}_{\rm{s}_\textit{q}} \mid {\rm{s}_\textit{q}} \right)$ represent the matching probability of sub-segment-scale salient features. In the (\ref{postpro}), the $Pr\left(\rm{s}_\textit{l} \mid \rm{r}_\textit{i} \right)$ denotes the prior probability of the sub-segment $\rm{s}_\textit{l}$ within the road $\rm{r}_\textit{i}$.  
\par
Based on the posterior probability $Pr\left(\rm{s}_\textit{l}\mid \hat{\boldsymbol{f}}_{\rm{s}_\textit{l}} \right)$, the vehicular road and sub-segment can be estimated through solving the following problem:  
\vspace{-2mm}
\begin{alignat}{2}	
	\label{segloc}
	\mathop{\max}_{i,l} \quad & \begin{array}{*{20}{c}}
		Pr\left(\rm{s}_\textit{l}\mid \hat{\boldsymbol{f}}_{\rm{s}_\textit{l}} \right)
	\end{array} & \\[0.01pt] 
	\mbox{s.t.}\quad
	& 1 \leq i \leq m, 1 \leq l \leq n_i, & \tag{\ref{segloc}a}
	\vspace{1mm}
\end{alignat}
where the constraint (\ref{segloc}a) gives the positioning space of roads and sub-segments. To tackle this problem, we propose a two-spatial-scale localization algorithm, where the road-scale salient features are matched to localize the vehicular road in the road space $\mathcal{R}$, and then the sub-segment-scale salient features are matched to position the vehicular sub-segment in the sub-segment space $\mathcal{S}_i$ within the localized road. The probability of the road-scale feature matching, i.e., $Pr\left(\hat{\boldsymbol{f}}_{\rm{r}_\textit{i}}\mid {r}_{i}\right)$ fits well to an exponential probability distribution \cite{newson2009hidden}, and can be expressed as:
\begin{equation}
	\setlength{\abovedisplayskip}{3pt}
	Pr\left(\hat{\boldsymbol{f}}_{\rm{r}_\textit{i}} \mid {\rm{r}_\textit{i}} \right) = e^{-{d\left(\hat{\boldsymbol{f}}_{\rm{r}_\textit{i}}, \boldsymbol{f}_{\rm{r}_\textit{i}}\right)}},
	\vspace{-1mm}
\end{equation}
where $d\left(\hat{\boldsymbol{f}}_{\rm{r}_\textit{i}}, \boldsymbol{f}_{\rm{r}_\textit{i}}\right)$ represents the $l_2$-norm between the vector of vehicular salient features and the vector of $j$th road-scale salient features in $\boldsymbol{f}_{\rm{r}_\textit{i}}$. The $d\left({\hat{\boldsymbol{f}}_{\rm{r}_\textit{i}}}, \boldsymbol{f}_{\rm{r}_\textit{i}}\right)$ can be obtained as follows:
\begin{equation}
	\setlength{\abovedisplayskip}{4pt}
	d\left({\hat{\boldsymbol{f}}_{\rm{r}_\textit{i}}}, \boldsymbol{f}_{\rm{r}_\textit{i}}\right) = \|\boldsymbol{W}_{\rm{r}_\textit{i}}\boldsymbol{e}_{u} - \boldsymbol{W}_{\rm{r}_\textit{i}}\boldsymbol{e}_{\rm{r}_\textit{i}}\|_2,
	\vspace{1mm}
\end{equation}
where $\boldsymbol{W}_{\rm{r}_\textit{i}}\boldsymbol{e}_{\rm{r}_\textit{i}}$ denotes the $j$th road's salient features in the road-scale salient feature vector $\boldsymbol{f_{\rm{r}}}$, and $\boldsymbol{W}_{\rm{r}_\textit{i}}\boldsymbol{e}_{u}$ represents the vehicular salient features corresponding to the $j$th road. The probability of sub-segment feature matching $Pr\left(\hat{\boldsymbol{f}}_{\rm{s}_\textit{l}} \mid \rm{s}_\textit{l} \right)$ can be given by
\begin{equation}
	\setlength{\abovedisplayskip}{2pt}
	Pr\left(\hat{\boldsymbol{f}}_{\rm{s}_\textit{l}} \mid \rm{s}_\textit{l} \right) =  e^{-{d\left({\hat{\boldsymbol{f}}_{\rm{s}_\textit{l}}}, \boldsymbol{f}_{\rm{s}_\textit{l}}\right)}},
	\vspace{-1mm}
\end{equation}
where $d\left({\hat{\boldsymbol{f}}_{\rm{s}_\textit{l}}}, \boldsymbol{f}_{\rm{s}_\textit{l}}\right)$ represents the $l_2$-norm between the vector of vehicular sub-segment-scale salient features and the vector of the $l$th sub-segment-scale salient features in $\boldsymbol{f_{\rm{s}}}$. The $d\left({\hat{\boldsymbol{f}}_{\rm{s}_\textit{l}}}, \boldsymbol{f}_{\rm{s}_\textit{l}}\right)$ can be expressed as follows:
\begin{equation}\label{segdis}
	\setlength{\abovedisplayskip}{2pt}
	d\left({\hat{\boldsymbol{f}}_{\rm{s}_\textit{l}}}, \boldsymbol{f}_{\rm{s}_\textit{l}}\right) = \|\boldsymbol{W}_{\rm{s}_\textit{l}}\boldsymbol{e}_{u} - \boldsymbol{W}_{\rm{s}_\textit{l}}\boldsymbol{e}_{\rm{s}_\textit{l}}\|_2,
	\vspace{1mm}
\end{equation}
where $\boldsymbol{W}_{\rm{s}_\textit{l}}\boldsymbol{e}_{u}$ represents the vehicular salient features corresponding to the $l$th sub-segment. The $\boldsymbol{W}_{\rm{s}_\textit{l}}\boldsymbol{e}_{\rm{s}_\textit{l}}$ in (\ref{segdis}) denotes the salient feature vector of the $l$th sub-segment in the $\boldsymbol{f_{\rm{s}}}$. Through the proposed two-spatial-scale localization algorithm, the vehicular roads and sub-segments can be quickly located with the maximum posterior probability.  
\par
To obtain fine-grained 2D coordinates within the localized sub-segment, the RSRP values of the sub-segment-scale HetNet signal sequence can be first mapped to coordinates through curve fitting in the offline stage. Then, the real-time latitude and longitude values of a mobile vehicle can be localized based on fitted curves.
\par
In the two-spatial-scale road localization process, the computation complexity is $\mathcal{O}\left(N_r+N_s+K\right)$, where $N_r$, $N_s$, and $K$ represent the number of all road-scale salient features, the number of all sub-segment-scale salient features within the localized road, and the number of HetNet BSs, respectively. Since sparse features, low-dimensional roads, and sub-segments are used in the localization process, the computation complexity is significantly reduced.
\section{Experiment Results and Analysis}
\label{Experiment Results}
In this section, we conduct experiments to evaluate the localization performance of the proposed road-aware localization mechanism in the HetNet system. Specifically, two performance metrics, i.e., the mean delay and the mean distance error (MDE) of located positions, were verified with different BS numbers and grid sizes. In the simulation, an area of $600$ m $\times$ $600$ m with a high-rise building around four roads is considered as the interested localization area. The number of feature kinds $q$ in the proposed mechanism is set as $5$. HetNet signals are sampled with $1$ m distance interval. The target vehicle moves at the velocity of $30$ Km/h. We have a full implementation for the proposed mechanism in Python 3.9 on a PC with 24 GB RAM.  
\par
Three approaches are implemented for performance comparison, i.e., the state of art restricted weighted k-nearest neighbor algorithm (RWKNN) \cite{RWKNN}, the gradient-based fingerprinting (GIFT) \cite{Gradient}, and the curve fitting-based exhaustive location search algorithm (CF-ELS) \cite{Wang2015Curve}. The former two respectively utilize the absolute value and the gradient feature of HetNet signals to locate vehicles through signal pattern matching. The CF-ELS is chosen to be compared as range estimations are used for localization. The number of nearest neighbors is set as 3 in RWKNN. The step size of the exhaustive search is set to 0.1 m \cite{Wang2015Curve}. To achieve effective experimental comparisons, a simulation data set and a real data set are employed to evaluate the localization performance. The simulation data set is collected through the Qualnet simulator \cite{Qualnet} based on real service parameters of HetNet BSs \cite{RECOMMENDATIONIP} through signal measurement reports. Each record of the simulation data set contains the RSRP values from 6 HetNet BSs and a 2D coordinate. By using an RF measurement tool (Cellular-Z) \cite{s20061691}, the real data set is collected, which records RSRP values from 2 HetNet BSs and a 2D coordinate on each signal sampling position.   
\begin{figure}
	\centering
	\scriptsize
	\begin{tabular}{cc}		
		\subfigure [Mean delay versus BS number.] {
			\vspace{4 mm}
			\label{figure:subfig_c}
			
			\centering
			\hspace{-6.5 mm}
			\includegraphics [width=4.8cm]{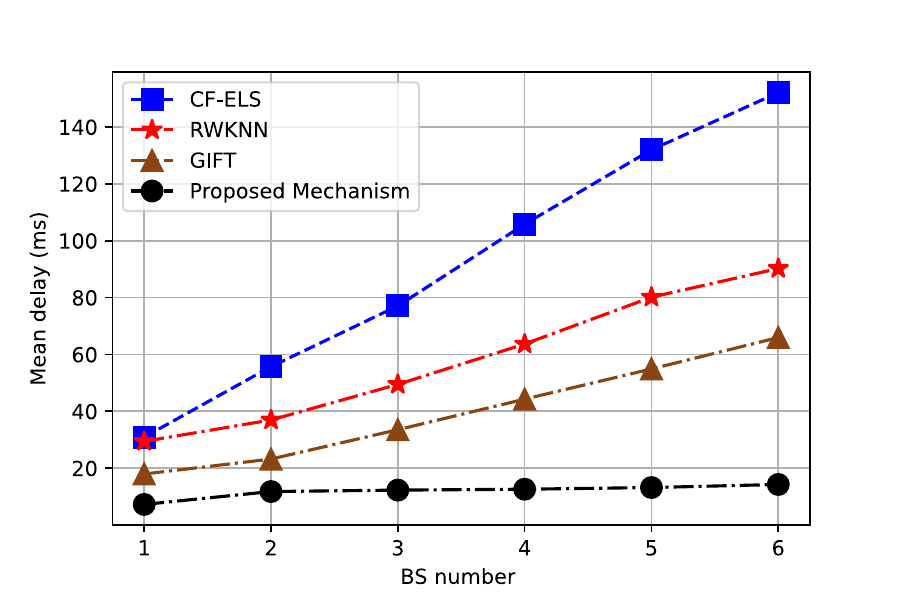}					
		} \hspace{-7mm}  		
		\subfigure [MDE versus BS number.] {
			\vspace{4 mm}
			\label{figure:subfig_d}
			\centering
			\includegraphics[width=4.8cm]{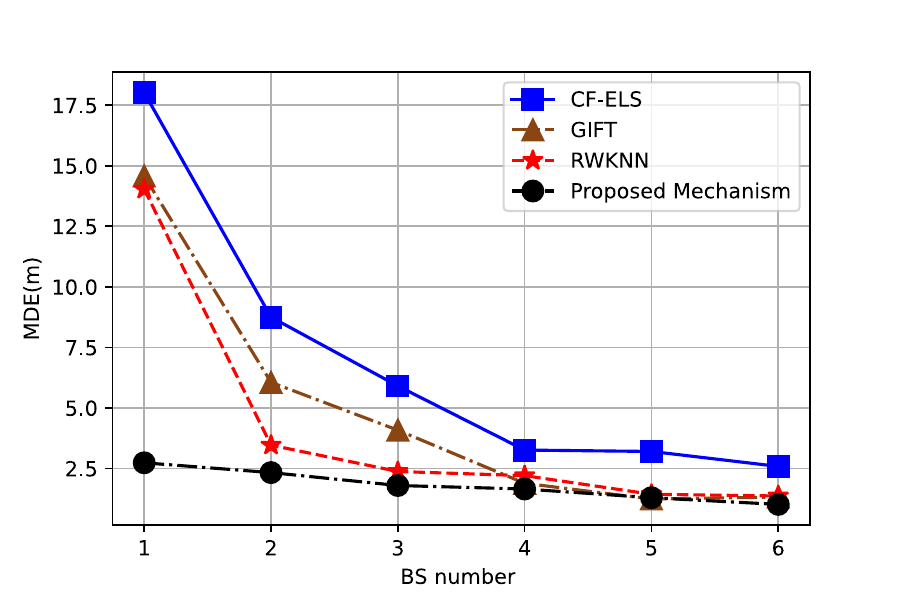}}\hspace{-5mm}		
		\centering
	\end{tabular}
	\vspace{-3 mm}
	\caption{Localization performance based on different BS numbers.}
	\label{figure65}
	\vspace{-5 mm}
\end{figure} 
\par
Based on the signals from 1 BS to 6 BSs, the mean delay and the MDE of positions within a real-time HetNet signal sequence are shown in Fig. \ref{figure:subfig_c} and Fig. \ref{figure:subfig_d}, respectively. In Fig. \ref{figure:subfig_c}, the proposed mechanism achieves the lowest mean delay compared with the benchmark method since salient signal features of HetNet BSs are utilized for the feature-matching localization. Fig. \ref{figure:subfig_d} demonstrates that by employing the two-spatial-scale localization algorithm, the MDE of the proposed mechanism outperforms benchmark methods under different BS numbers. Particularly, the proposed mechanism can achieve the lowest MDE of $2.43$ m when signals of 2 BSs are received by VUE due to the significant signal attenuation in the 5G HetNet. To sum up, based on different BS signals, the proposed mechanism can localize UE with the lowest localization error and the lowest localization latency compared with the RWKNN, GIFT, and CF-ELS.
\begin{figure}
	\centering
	\scriptsize
	\begin{tabular}{cc}		
		\subfigure [Mean delay versus grid size.] {
			\vspace{4 mm}
			\label{figure:subfig_e}
			
			\centering
			\hspace{-6.5 mm}
			\includegraphics [width=4.8cm]{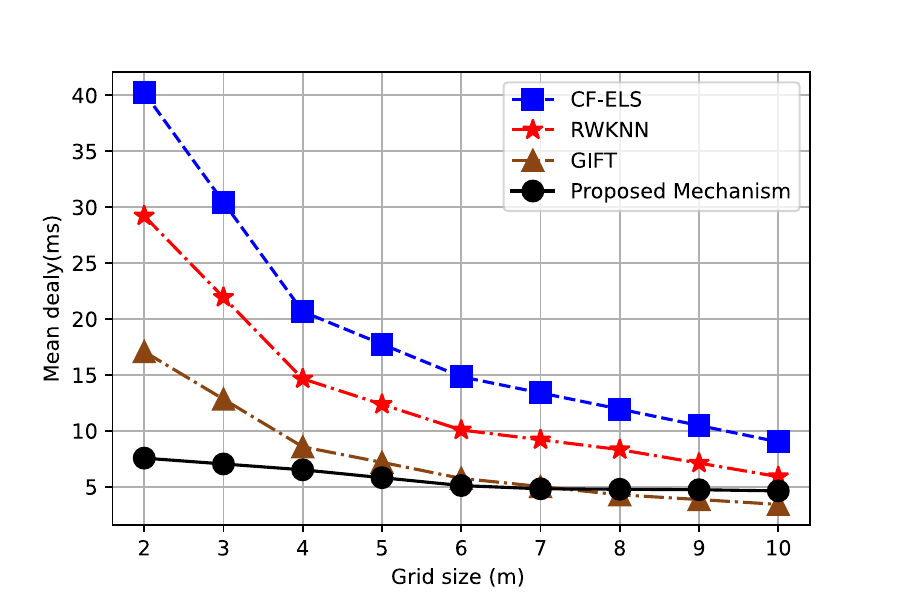}					
		} \hspace{-7mm}  		
		\subfigure [MDE versus grid size.] {
			\vspace{4 mm}
			\label{figure:subfig_f}
			\centering
			\includegraphics[width=4.8cm]{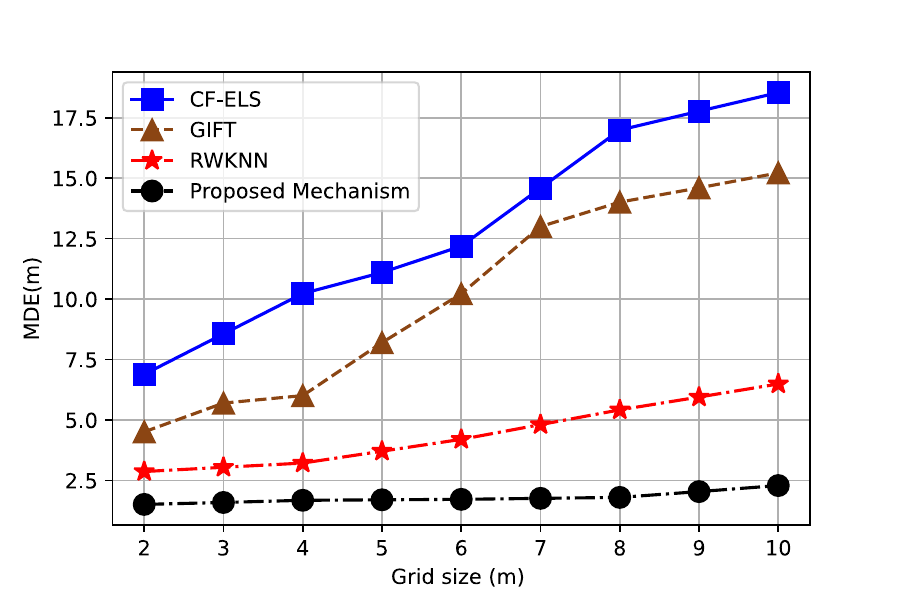}}\hspace{-5mm}		
		\centering
	\end{tabular}
	\vspace{-1 mm}
	\caption{Localization performance at different grid sizes.}
	\label{figure87}
	\vspace{-1 mm}
\end{figure}
\par
Fig. \ref{figure:subfig_e} and Fig. \ref{figure:subfig_f} show the localization performance under different grid sizes. As shown in Fig.\ref{figure:subfig_e}, the lowest mean delay is achieved by the proposed mechanism, measuring less than $10$ ms. This is due to the fact that roads and sub-segments are divided according to the actual length of the road and HetNet signal features, thereby avoiding reliance on uniform grid sizes. The MDE of the proposed algorithm in Fig. \ref{figure:subfig_f} is below $2.5$ m as the grid size increases. However, though the mean delay of benchmark methods decreases, the MDEs increase on the larger grid size due to coarse-grained fingerprints. These results indicate that the proposed mechanism can achieve low-latency vehicle localization under large grid sizes, which can consume less human cost and storage space to maintain two-spatial-scale salient features.
\par
\begin{figure}[t]
	\centering
	\includegraphics[width=7.5cm]{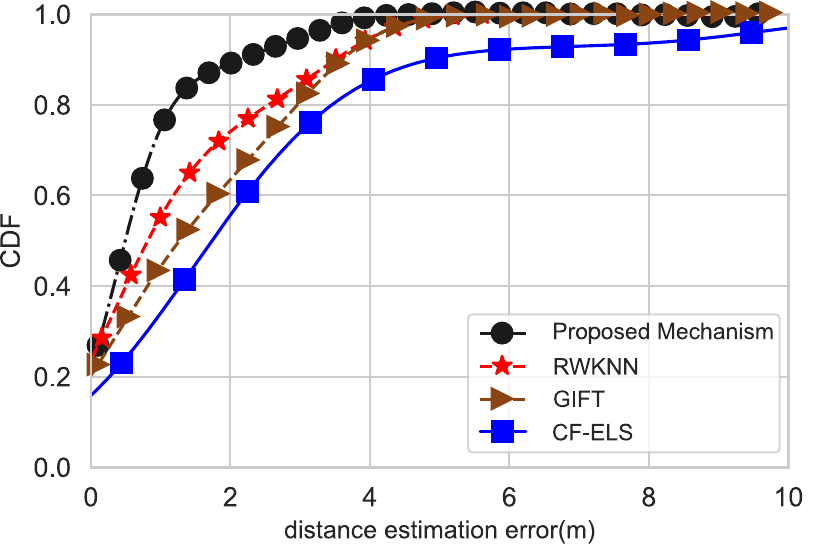}
	\vspace{-1mm}
	\caption{CDF of localization error based on real signal measurements.}
	\label{figure9}
	\vspace{-4mm}
\end{figure}

Fig. \ref{figure9} illustrates the cumulative distribution function (CDF) of localization errors based on the real data set at the $2$ m grid size. It can be observed from Fig. \ref{figure9} that the proposed algorithm has respectively $17\%$, $27\%$, $35\%$ improvement (from 2 m to 0 m) compared with the RWKNN, the GIFT, and the CF-ELS.
\par
Table \ref{locationtime} shows the mean delay of the four localization methods based on the real data set. Since the position search spaces are partitioned into road scale and sub-segmented scale, the mean delay of our proposed method could be significantly reduced to $7.2$ ms. For high-speed vehicles, the localization delay below 10 ms is essential for keeping a safe distance between vehicles \cite{SelfDriving}. Since the localization times of the RWKNN, the GIFT, and the CF-ELS are more than 10 ms, it is hard to fulfill the vehicle localization requirements with low latency and high reliability.
\newcommand{\tabincell}[2]{\begin{tabular}{@{}#1@{}}#2\end{tabular}}
\begin{table}[h]	
	\centering	
	\vspace{-2mm}
	\caption{Mean delay of located positions}
	\vspace{-1mm}
	\begin{tabular}{|c|c|c|c|c| }
		\hline 
		Method & {RWKNN} & GIFT & {CF-ELS} & \tabincell{c} {Proposed \\ Mechanism}\\
		\hline 
		\tabincell{c} {Mean delay} & 32.7 ms & 28.5 ms & 57.8 ms & 7.2 ms \\
		\hline	 
	\end{tabular}
	\vspace{-1mm}
	\label{locationtime}
\end{table}
\section{Conclusions}
\label{Conclusions}
In this paper, we proposed a road-aware localization mechanism for low-latency vehicle localization. Specifically, a HetNet signal sequence segmentation method and a salient feature extraction method were devised to sparsely represent roads and sub-segments. Then, a two-spatial-scale vehicular localization algorithm was designed through road and sub-segment salient feature matching. Fine-grained positioning was achieved through curve fitting to localize vehicular coordinates. The numerical results validated that the proposed mechanism provided a promising solution to reduce localization latency and support reliable localization accuracy. The HetNet system with multiple MBSs and SBSs will be further studied in our future work.

\footnotesize
\bibliography{mybibfile}

\begin{thebibliography}{10}

\bibitem{8827665}
K.-H. Lam, C.-C. Cheung, and W.-C. Lee, ``{RSSI}-based {LoRa} localization
  systems for large-scale indoor and outdoor environments,'' {\em IEEE Trans.
  Veh. Technol.}, vol.~68, pp.~11778--11791, Sep. 2019.

\bibitem{requirement}
{EUSPA}, ``The earth observation ({EO}) and global navigation satellite system
  ({GNSS}) market report.'' [Online], Jan. 2022.
\newblock
  \url{https://space-economy.esa.int/article/126/euspa-publishes-eo-and-gnss-market-report-2022}.

\bibitem{Jian2018MEMS}
K.~Jian, N.~Xiaoji, and C.~Xingeng, ``Robust pedestrian dead reckoning based on
  {MEMS} {IMU} for smartphones,'' {\em IEEE Sens. J.}, vol.~18, p.~1391–1409,
  May 2018.

\bibitem{Zakaria2017PerformanceEO}
Y.~Zakaria and L.~Ivanek, ``Performance evaluation of {UE} location techniques
  in {LTE} networks,'' {\em Am. J. Sci.}, vol.~14, pp.~81--89, Jan. 2017.

\bibitem{4510769}
K.~Yu and Y.~J. Guo, ``Statistical {NLOS} identification based on {AOA}, {TOA},
  and signal strength,'' {\em IEEE Trans. Veh. Technol.}, vol.~58,
  pp.~274--286, Jan. 2009.

\bibitem{2023Kaitao}
K.~Meng, Q.~Wu, S.~Ma, W.~Chen, K.~Wang, and J.~Li, ``Throughput maximization
  for {UAV-enabled} integrated periodic sensing and communication,'' {\em IEEE
  Trans. Wireless Commun.}, vol.~22, pp.~671--687, Jan. 2023.

\bibitem{Abbas2018NLOS}
A.~Abolfathi~Momtaz, F.~Behnia, R.~Amiri, and F.~Marvasti, ``{NLOS}
  identification in range-based source localization: Statistical approach,''
  {\em IEEE Sens. J.}, vol.~18, pp.~3745--3751, May 2018.

\bibitem{Giovanni2018Passive}
G.~Pecoraro, E.~Cianca, S.~Di~Domenico, and M.~De~Sanctis, ``{LTE} signal
  fingerprinting device-free passive localization robust to environment
  changes,'' in {\em Proc. Glob. Wireless Summit (GWS)}, pp.~114--118, Nov.
  2018.

\bibitem{2017HetNet}
P.-H. Tseng and K.-T. Lee, ``A femto-aided location tracking algorithm in
  {LTE-A} heterogeneous networks,'' {\em IEEE Trans. Veh. Technol.}, vol.~66,
  no.~1, pp.~748--762, 2017.

\bibitem{6821742}
F.~Zhao, H.~Luo, H.~Geng, and Q.~Sun, ``An {RSSI} gradient-based {AP}
  localization algorithm,'' {\em China Commun.}, vol.~11, pp.~100--108, Feb.
  2014.

\bibitem{5462020}
A.~K. M.~M. Hossain and W.-S. Soh, ``Cramer-rao bound analysis of localization
  using signal strength difference as location fingerprint,'' in {\em IEEE
  INFOCOM}, pp.~1--9, Mar. 2010.

\bibitem{TRUONG2020}
C.~Truong, L.~Oudre, and N.~Vayatis, ``Selective review of offline change point
  detection methods,'' {\em Signal Processing}, vol.~167, p.~107299, 2020.

\bibitem{newson2009hidden}
P.~Newson and J.~Krumm, ``Hidden markov map matching through noise and
  sparseness,'' in {\em Proc. ACM SIGSPATIAL}, Nov. 2009.

\bibitem{RWKNN}
G.~Chen, X.~Guo, K.~Liu, X.~Li, and J.~Yang, ``{RWKNN}: A modified {WKNN}
  algorithm specific for the indoor localization problem,'' {\em IEEE Sens.
  J.}, vol.~22, pp.~7258--7266, Apr. 2022.

\bibitem{Gradient}
Y.~Shu, Y.~Huang, J.~Zhang, {\em et~al.}, ``Gradient-based fingerprinting for
  indoor localization and tracking,'' {\em IEEE Trans. Ind. Electron.},
  vol.~63, pp.~2424--2433, Apr. 2016.

\bibitem{Wang2015Curve}
B.~Wang, S.~Zhou, W.~Liu, and Y.~Mo, ``Indoor localization based on curve
  fitting and location search using received signal strength,'' {\em IEEE
  Trans. Ind. Electron.}, vol.~62, pp.~572--582, Jan. 2015.

\bibitem{Qualnet}
``Qualnet network simulator software.''
  \url{https://www.ncs-in.com/product/qualnet-network-simulator-software/}.

\bibitem{RECOMMENDATIONIP}
``Recommendation {ITU-R P.1411-11} - propagation data and prediction methods
  for the planning of short-range outdoor radiocommunication systems and radio
  local area networks in the frequency range 300 {MHz} to 100 {GHz},'' Geneva:
  Electronic Publication, Sep. 2021.

\bibitem{s20061691}
D.~Li, Y.~Lei, and H.~Zhang, ``A novel outdoor positioning technique using
  {LTE} network fingerprints,'' {\em Sensors}, vol.~20, Mar. 2020.

\bibitem{SelfDriving}
B.-J. Chang, W.~Hung, Y.~Lin, and W.-T. Chang, ``Dynamic keeping reserved
  resource probability with slicing flow steering in {5G} sidelink {SPS} for
  platooning {ADAS} and autonomous self driving,'' in {\em Proc. Int. Autom.
  Control Conf. (CACS)}, pp.~1--6, Dec. 2020.

\end{thebibliography}
\bibliographystyle{ieeetr}

\clearpage
\normalsize

\end{document}